\documentclass[aps,10pt,pre,floatfix,twocolumn,superscriptaddress]{revtex4-1}
\usepackage{color}
\usepackage{ulem}
\usepackage{graphicx}
\usepackage{subfigure,amsmath, amsthm, amssymb}
\usepackage[colorlinks,citecolor=red]{hyperref}

\newcommand{\be}{\begin{equation}}
\newcommand{\ee}{\end{equation}}
\newcommand{\bea}{\begin{eqnarray}}
\newcommand{\eea}{\end{eqnarray}}

\begin{document}
\title{Nucleation rate in the two dimensional Ising model in the presence of random impurities}
\author{Dipanjan Mandal}
\email{dipanjan.mandal@warwick.ac.uk}
\affiliation{Department of Physics, University of Warwick, Coventry CV4 7AL, United Kingdom}
\author{David Quigley}
\email{d.quigley@warwick.ac.uk}
\affiliation{Department of Physics, University of Warwick, Coventry CV4 7AL, United Kingdom}

\date{\today}
\begin{abstract}
Nucleation phenomena are ubiquitous in nature and the presence of impurities in every real and experimental system is unavoidable. Yet numerical studies of nucleation are nearly always conducted for entirely pure systems. We have studied the behaviour of the droplet free energy in two dimensional Ising model in the presence of randomly positioned static and dynamic impurities. We have shown that both the free energy barrier height and critical nucleus size monotonically decreases with increasing the impurity density for the static case. 
We have compared the nucleation rates obtained from Classical Nucleation Theory and the Forward Flux Sampling method for different densities of the static impurities. The results show good agreement. In the case of dynamic impurities, we observe preferential occupancy the impurities at the boundary positions of the nucleus when the temperature is low. This further boosts enhancement of the nucleation rate due to lowering of the effective interfacial free energy.
\end{abstract}
\maketitle

\section{Introduction}
\label{intro}
Nucleation phenomena are of great importance both in physics and chemistry. The study of nucleation has a long history, rooted in the widely used theoretical model of classical nucleation theory (CNT)~\cite{Kashchiev2000} and the Becker-Doring (BD) expression for the nucleation rate~\cite{1935-becker-aphys}. CNT is able to explain the qualitative (and sometimes quantitative) behaviour, of various nucleation processes occurring in natural and experimental systems~\cite{villuendas-2007-prl,peter-1992-metastable_water}. The nucleation rate at which a stable phase nucleates from a metastable parent depends on various parameters, e.g., temperature~\cite{pitto-2019-temperature}, pressure~\cite{1992-lee-jap}, external electric field~\cite{kashchiev-1972-electric-field}, viscosity~\cite{frank-2007-viscosity}, application of shear~\cite{allen2008jcp}, etc. It is also heavily influenced by  surfaces and the presence of impurities~\cite{rajiv-1993-impurities,keshavarz-2019-impurity} in the system. 

Modelling nucleation from solution is an active area of research with relevance to pharmaceutical manufacture, biomineral formation and other highly complex precipitation processes. Atomistic simulations~\cite{agarwal_solute_2014} are now able to make quantitative predictions of the nucleation rate. Unfortunately, even for nucleation of simple salt crystals from solution~\cite{jiang_forward_2018,zimmermann_nucleation_2015} agreement with experiment remains elusive and sensitive to details of the atomistic force field and the order parameter used to identify solid regions in the simulation~\cite{doi:10.1063/1.5024009}. In addition, the solution is nearly always simulated as pure with no impurities or spectator/counter ions. A quantitative understanding of how these additional species impact on the nucleation rate is lacking. 

The most basic physics of nucleation from solution can be captured using a simple lattice-based model of solute precipitation~\cite{binder_jcp_2016}. Minimal models of this kind in two and three dimensions have been used to probe assumptions of CNT and to test advanced methods for quantifying rare events against well-established results~\cite{ford1999pre,cai2010pre,1994-sides-pre,lifanov_jcp_2016,dietrich-1982-prl}. Extensions of the basic Ising-like model have been used to study nucleation inside a rectangular pore~\cite{page2006prl,whitelam2012sm}, on substrates~\cite{binder_hetero_2018}, two-step nucleation mechanisms \cite{duff_jcp_2009} and nucleation in the presence of shear~\cite{allen2008jcp}.  Studies have also explored the choice of microscopic kinetics on the nucleation rate~\cite{kuipers_limitations_2010}. In this work we study such a model in the presence of neutral impurities as a further step toward introducing some of the complexity found in ``real'' solutions. We explore how this changes nucleation rate, critical nucleus size and free energy barrier height, as a function of impurity concentration. 

We consider two scenarios, static and dynamic impurities. In the static impurity case the impurities are sites distributed randomly in space and fixed throughout the simulation. We work with quantities averaged over many samples of this static disorder. In the dynamic case impurities can migrate via Kawasaki dynamics~\cite{kawasaki-1966}. A particular advantage of working with such simplified models is that the timescale of impurity migration relative to cluster growth can be adjusted to probe different kinetic regimes. 

We have used umbrella sampling techniques to calculate the free energy barrier for different densities of the impurity sites. We also calculate the nucleation rate using the Forward flux sampling (FFS) technique and compare the results to those computed from the free energy barrier via the Becker-Doring expression. We have shown that the impurity particles preferentially occupy the boundary positions of the nucleus for the dynamic case when the temperature is low. This surface accumulation process enhances the nucleation rate by a multiple of $10^4$ compared to the static case.

The rest of this paper is organised as follows. In Sec.~\ref{model}, we describe the model and  simulation techniques used to study the model. Variation of the free energy barrier height and the critical nucleus size are shown in Sec.~\ref{barrier}. Section~\ref{nucl-rate} contains the comparison of theoretical and computational nucleation rates for different impurity concentration. The effect of dynamic impurities on barrier height as well as nucleation rates are described in Sec.~\ref{dynamic-imp}. Finally we conclude in Sec.~\ref{conclusion}.


\section{Model \& simulation techniques}
\label{model}
We consider the two dimensional Ising model on a $L\times L$ square lattice with spin $S_i$ at position $i$. The Hamiltonian of the model may be written as
\be
\mathcal{H}=-J\sum_{\langle i,j\rangle}S_{i}S_{j}-h\sum_{i}S_{i},
\ee
where $J$ is the strength of the coupling between two nearest neighbours, $h$ is the externally applied field and ${\langle i,j\rangle}$ represents summation over all nearest-neighbour pairs in the usual manner. The spin can take two values, $S_i = \pm 1$. We take the strength of the coupling to be positive ($J>0$), i.e., the ferromagnetic Ising model. This is analogous to a lattice-gas model in which each lattice site can be occupied ($S_i=+1$) or unoccupied ($S_i=-1$), with the external field playing the role of a particle reservoir at a constant chemical potential. 

We study the discontinuous magnetisation reversal for a system with initially negative magnetisation in the presence of a positive magnetic field. This necessarily occurs only for temperatures $T<T_{c}$ ($T_c$ is the critical temperature) and proceeds via nucleation and growth of domains with positive magnetisation. In the lattice-gas (solute precipitation) analogy this corresponds to an initially mostly-empty lattice (solvent-rich) to a full lattice (solute-rich) at a chemical potential where equilibrium with a particle reservoir corresponds to a mostly-occupied lattice. 

We modify the model by introducing a third spin state $S_i=0$ at random positions of the lattice. This represents non-magnetic impurities, or ``neutral'' impurity ions in solution which do not favour interaction with either solute or solvent neighbours. Non-zero spins are evolved in time via the usual  spin-flip dynamics in which a randomly selected spin is flipped, and the move accepted or rejected according to the Metropolis criterion. In what follows we mainly use the magnetic terminology for brevity.

The overall density of spin-$0$ impurities is kept fixed throughout the simulation. Further, we consider two different circumstances, static and dynamic impurities. In the static case, the impurity particles are immobile. In the dynamic case impurities can diffuse through the lattice using spin-exchange dynamics often known as Kawasaki dynamics~\cite{kawasaki-1966}. The ratio of impurity-diffusion move attempts to spin flip move attempts controls the mobility of impurities relative to the nucleation timescale.

\begin{figure}
\includegraphics[width=\columnwidth]{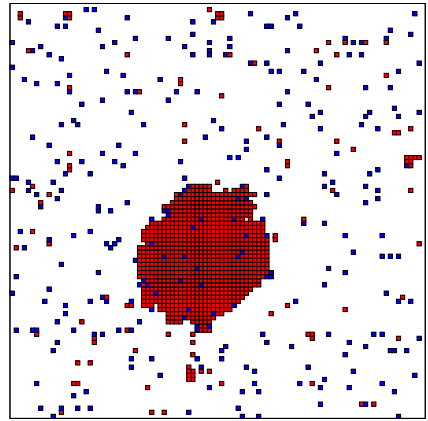}
\caption{Typical configuration of the  nucleating system in the presence of static impurities for $T=1.5$, $h=0.05$, $\rho_i=0.028$ and $L=100$. Particles and impurities are respectively denoted by red and blue colors respectively, and white represents empty site.}
\label{fig:snapshot}
\end{figure}

We calculate the nucleation rate using two independent methods, and as a function of the impurity density. The first method calculates the free energy barrier to nucleation and applies the BD expression for the nucleation rate and the second method is the Forward Flux Sampling (FFS) method which calculates the rate independently of classical nucleation theory. A typical configuration of the system during the transition is shown in Fig.~\ref{fig:snapshot} where red represents occupied ($S_i=1$) sites, blue represents impurity ($S_{i}=0$) and white represents empty sites ($S_i=-1$). 

In the next sections, we describe in more detail the simulation techniques used to calculate the barrier heights and the nucleation rates. 

\subsection{Umbrella Sampling}

In situations where the free energy barrier to nucleation is large,  it is impossible to adequately sample the probability distribution for large nuclei in an unbiased simulation. Umbrella sampling (US) is widely used biased simulation technique to calculate  free energy as a function of a collective variable or  reaction coordinate. The US method was introduced by Torrie and Valleau in 1976~\cite{us_torrie_1977} and tested successfully for Lennard-Jones systems. The method has been used in different contexts, e.g., protein folding dynamics in biological systems, measuring reaction rates in chemical systems, etc~\cite{charles-2018-us}. The method is also often used (as here) in the context of nucleation~\cite{frenkel_2004_spherical_colloids,us_review_2011} to obtain free energy as a function of cluster size and hence parameterise a BD calculation of the nucleation rate. We define a cluster as a contiguously connected region of $+1$ spins. This definition has proven to yield accurate BD estimates of the nucleation rate in the regime of interest here~\cite{cai2010pre}, but may not be optimal closer to $T_{c}$~\cite{schmitz_tests_2013}. The configuration space is divided into overlapping windows of equal size which span the range of cluster sizes from one to the nucleus size greater than the critical nucleus size. The probability distribution of the cluster sizes lying inside the window range is calculated for each window. In our US simulations all windows  span a cluster size of range $20$. A range of $10$ overlaps between two neighbouring windows. We have taken wide overlap between two windows to reduce the error while combining different parts of the free energies. We use an infinite square well umbrella potential of the same width as the window. For each window we run the simulation for $10^9$ steps where each step is an attempt to flip a randomly chosen spin. We skip the spin update if the the random site contain an impurity spin. After each step we measure cluster sizes. If the largest cluster size escapes the window range we reject the spin flip and restore the previous state of the spin. Simulations of this kind are particularly suited to parallelisation over windows, for which we employ OpenMP~\cite{openmp_1998} threading to decrease simulation time significantly. In the case of static impurities, we average the obtained probability distribution for each window with $28$ different impurity configurations. An uncertainty in the probability distribution is obtained by averaging over $28$ independent calculations of probability distribution in each window. The maximum error is obtained from the standard error $\sigma/\sqrt{\mathcal{N}}$ of the probability at each nucleus size, where $\sigma$ is the standard deviation of $\mathcal{N}$ independent results.

The relative free energy for a droplet of size $\lambda$ at the $n$-th window may be written as
\be
f^{US}_n(\lambda)=-k_BT\ln[P_n(\lambda)],
\label{eq:define_free_en}
\ee
where $P_n(\lambda)$ is the probability of sampling a cluster of size $\lambda$ and $k_B$ is the Boltzmann constant. We take $k_B=1$ in our calculations. We combine the free energy obtained for the $n$-th window with the free energies of the previous windows by a constant shift $S_n$ which may be written as
\be
S_n=\frac{1}{m}\sum_i-k_BT\ln\bigg[\frac{P_{n-1}(i)}{P_n(i)}\bigg],
\ee
where $m$ is the total number of overlapped points and $i$ runs over all overlapped points between $(n-1)$-th and $n$-th windows. After the constant shift the free energy of the $n$-th window may be written as
\be
F^{US}_n(\lambda)=f^{US}_n(\lambda)+S_n.
\ee
Once all windows have been combined in this fashion we obtain $F^{US}(\lambda)$, the free energy over the entire sampled range of cluster sizes.

\subsection{Forward Flux Sampling}
FFS~\cite{2009_ffs_allen,2009_tps_escobedo,2005_ffs_allen,polymer_folding_allen_2012} is a numerical technique widely used to calculate the rate of occurrence of a rare event. In this paper we calculate the nucleation rate from the initial metastable phase (negative spin) to the stable phase (positive spin) using FFS. The method proceeds via the following steps. First, we divide the configuration space connecting the metastable and the stable phase with intermediate equally spaced interfaces $\lambda_{i}$. Each interface is an isosurface of the maximum cluster size, such that progression from the first interface $\lambda_{0}$ to the final interface $\lambda_{N}$ captures a nucleation trajectory.  We assume that $\lambda_{i+1}$ is always greater then $\lambda_i$.  
\begin{figure}
\includegraphics[width=\columnwidth]{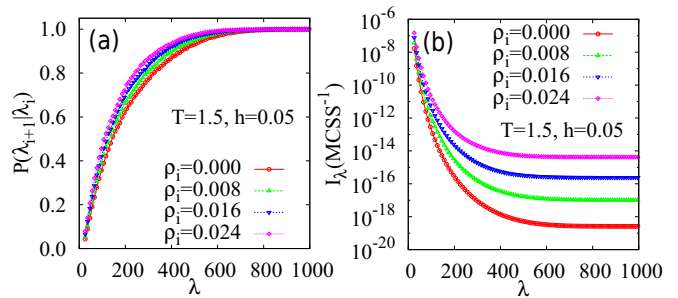}
\caption{Plots of (a) interface probability $P(\lambda_{i+1}|\lambda_i)$ and (b) nucleation rate $I_\lambda$ to a cluster of size $\lambda$ as a function of $\lambda$ for different impurity densities $\rho_i$ with fix temperature $T=1.5$ and field  $h=0.05$.}
\label{fig:transition_prob}
\end{figure}
The nucleation rate is then estimated as
\be
I^{FFS}=I_0\prod_{i=0}^NP(\lambda_{i+1}|\lambda_i),
\label{eq:ffs_rate}
\ee
where $I_{0}$ is the initial positive flux through interface $\lambda_0$, i.e., the number of crossings of $\lambda_{0}$ per unit area in unit time. We define a unit of time as $L\times L$ attempts to flip a single spin. The units of $I_{0}$ and hence $I^{FFS}$ are therefore crossings per Monte Carlo step per single site, or $MCSS^{-1}$.  $P(\lambda_{i+1}|\lambda_i)$ is the probability that a sequence of Monte Carlo moves beginning with a configuration with maximum cluster size $\lambda_i$ will cross to $\lambda_{i+1}$ before reaching $\lambda_A = 8$, i.e., reaches the next interface before returning to the metastable parent phase ($\lambda \leq\lambda_A$). 

In our simulations, we have used $250$ interfaces with a constant gap of $10$ between adjacent interfaces. The position of the first interface varies for different temperatures. For $T=1.5$ and $h=0.05$, the first interface position is $\lambda_0=16$. The position of first interface is chosen such that crossings are infrequent, i.e. the probability of being at $\lambda=\lambda_0$ state is quite low ($\lambda_0$ is above the 95th percentile of the cluster probability in the parent phase). We run  trials starting from the metastable (negative spin) phase to calculate the initial positive flux, storing configurations which cross $\lambda_0$ in the direction of increasing $\lambda$. We then initialise Monte Carlo trajectories from randomly chosen configurations at this interface and count how many reach $\lambda_1$ before returning to $\lambda_A$ to obtain the crossing probability of the first interface. This is repeated for all subsequent interface pairs. In each case we sample the interface until $25,200$ successful events are captured. The initial flux is calculated based on how much time is needed to produce $25,200$ positive crossings of $\lambda_0$. In a typical FFS simulation we sample interface up to $\lambda \sim 2500$. Plots of the interface probabilities at $\lambda=\lambda_i$ and the rate of reaching state $\lambda$ starting from the metastable phase for different impurity densities $\rho_i$ are shown in Fig.~\ref{fig:transition_prob}(a) and Fig.~\ref{fig:transition_prob}(b) respectively for fixed $T=1.5$ and $h=0.05$. The interface probability and rate saturate for large $\lambda$ indicating post-critical behaviour. The saturation becomes slower with increasing $\rho_i$.

\section{Barrier Height and the critical nucleus size}
\label{barrier}

We first report on variation of the nucleation behaviour in the presence of randomly distributed static impurity spins.

\begin{figure}[t!]
\includegraphics[width=\columnwidth]{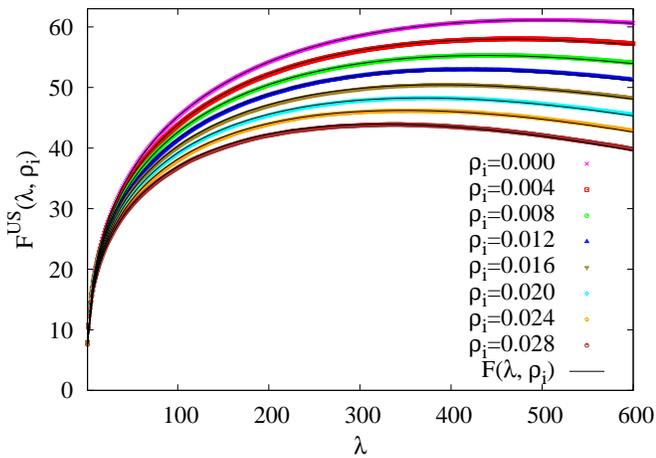}
\caption{Plot of the free energy $F^{US}(\lambda, \rho_i)$ obtained from the US simulations as a function of the nucleus size $\lambda$ for different static impurity densities $\rho_i$ at $T=1.5$ and $h=0.05$. The black line is the fit to the modified CNT expression of the free energy $F(\lambda, \rho_i)$ (see Eq.~\ref{eq:f_lambda_rho}).}
\label{fig:free_energy}
\end{figure}
The free energy of a cluster of $+1$ spins of size $\lambda$ may be written as
\be
F^{US}(\lambda)=-k_BT\ln\rho_1-k_BT\ln\bigg[\frac{P(\lambda)}{P(1)}\bigg],
\ee
where $\rho_1$ is the monomer density, i.e., the fraction of the sites occupied by isolated positive spins in the metastable parent phase and $P(\lambda)$ is the probability that an observed cluster will be of size $\lambda$. For a cluster size upper limit of $\lambda=\lambda_m$, this probability obeys the normalisation condition
\be
\sum_{\lambda=1}^{\lambda_m}P(\lambda)=1.
\ee
The variation of the free energy $F^{US}(\lambda)$ for different densities of the impurities $\rho_i$ is shown in Fig.~\ref{fig:free_energy}. The height of the free energy barrier decreases monotonically with increasing $\rho_i$. The nucleus size at which the barrier height takes maximum value is known as the critical nucleus size $\lambda_c$ which monotonically decreases with increasing $\rho_i$. This suggests that the presence of spin-0 impurities enhances nucleation of $S_{i}=+1$ domains.

\begin{figure}[t!]
\includegraphics[width=\columnwidth]{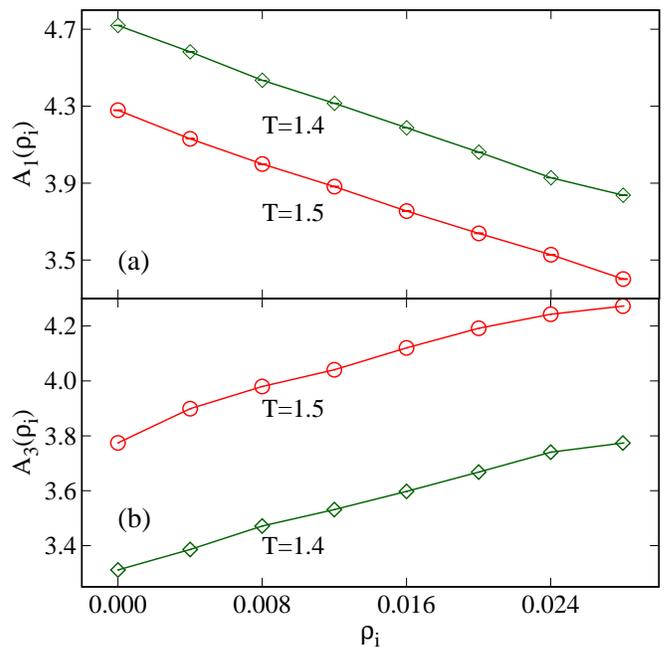}
\caption{Variation of (a) the interfacial tension $A_1(\rho_i)$, and (b) the size-independent term $A_3(\rho_i)$ as a function of the impurity densities $\rho_i$ for $T=1.5$ and $h=0.05$. The maximum impurity density is $\rho_i=0.028$ i.e. $2.8\%$ of the lattice is occupied by the impurities.}
\label{fig:a3_a1}
\end{figure}

To gain insight into this enhancement, we fit our free energy barriers to nucleation theory and extract trends in the resulting physical parameters. 
In a previous study~\citeauthor{cai2010pre}~\cite{cai2010pre} have numerically calculated free energy barriers in the absence of impurities and confirmed these accurately obey a cluster free energy given by
\be
F(\lambda)=-2h\lambda+A_1 \sqrt{\lambda}+A_2 \ln(\lambda)+A_3,
\label{eq:f_lambda}
\ee
where $A_1$ is proportional to the interfacial tension per unit length, $A_2$ is the coefficient of the logarithmic correction term introduced by Langer~\cite{1968-langer-prl} to incorporate the microscopic fluctuation of the cluster shape. In Ref.~\cite{cai2010pre}, the term $A_3$ is obtained from matching $F(\lambda)$ with the analytical expression of droplet free energy~\cite{shneidman-1999-jcp} for small $\lambda$. However, in the presence of impurities, we have used modified expression of free energy which can be written as
\bea
\label{eq:f_lambda_rho}
F(\lambda,\rho_i)&=&-2h\lambda+A_1(\rho_i) \sqrt{\lambda}+A_2 \ln(\lambda)+A_3(\rho_i),\\
\label{eq:A3}
A_3(\rho_i)&=&-k_BT\ln{\rho_1}-A_1(\rho_i)+2h.
\eea
The expression of $A_3(\rho_i)$ is obtained from the equality $F(1,\rho_i)=-k_BT\ln\rho_1$, where $\rho_1$ is the density of isolated $+1$ spins in the presence of impurities.
The term $A_2$ can be calculated theoretically for homogeneous nucleation and written as $A_2=\frac{5}{4}k_BT$~\cite{jacucci_1983}. We fit the free energy plot of the system without impurities obtained from the umbrella sampling techniques to the function $F(\lambda,\rho_i)$ written in Eq.(\ref{eq:f_lambda_rho}) to estimate $A_1(\rho_i)$ and $A_3(\rho_i)$. The fitted values of the parameters $A_1(\rho_i)$ and $A_3(\rho_i)$ are respectively $4.279$ and $3.776$ for $T=1.5$ and $h=0.05$, which are in very close agreement with $A_1\approx4.3$ and $A_3\approx3.7$, values obtained in reference~\cite{cai2010pre} for $\rho_i=0$.

\begin{figure}[t!]
\includegraphics[width=\columnwidth]{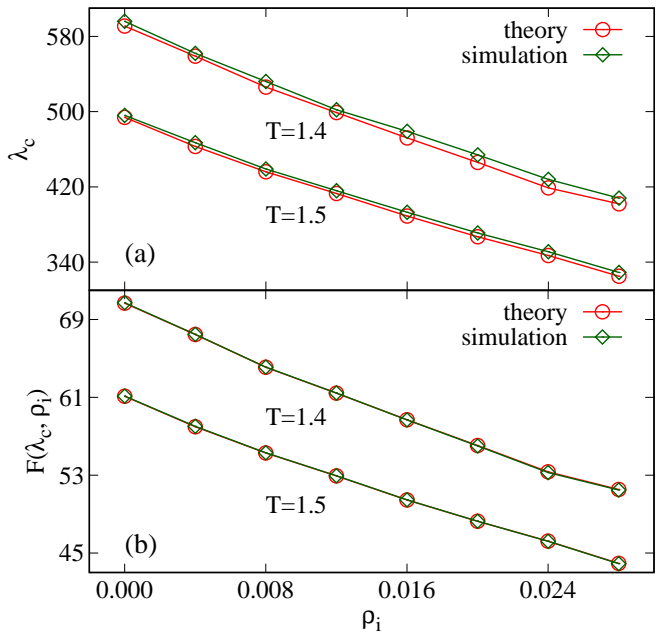}
\caption{Variation of (a) the critical nucleus size $\lambda_c$ and (b) the maximum free energy barrier $F(\lambda_c, \rho_i)$ as a function of $\rho_i$ for $T=1.5$ and $h=0.05$.}
\label{fig:lambdac_fc}
\end{figure}

In the presence of impurities we fix $A_2=\frac{5}{4}k_BT$ for all $\rho_i$. Allowing this parameter to vary has negligible impact on the quality of fit suggesting that the low impurity densities we study here do not significantly alter fluctuations in cluster shape.

The size-independent term $A_3(\rho_i)$ is obtained from the density of isolated $+1$ spins during unbiased simulations of the system in the metastable phase. This varies with impurity concentration. The only  free parameter when fitting the free energy plots in Fig.~\ref{fig:free_energy} for different $\rho_i$  is the interfacial tension per unit length $A_1(\rho_i)$. As shown in Fig.~\ref{fig:a3_a1}(a), this decreases linearly with increasing  impurity density for fixed $T$ and hence can be written as
\be
A_1(\rho_i)=m_1\rho_i+m_2.
\ee
It is interesting to explore how the parameters $m_1$ and $m_2$ vary with temperature and field. We observe that $A_1(\rho_i)$ increases with decreasing $T$ for fixed $\rho_i$. The variation of $A_3(\rho_i)$ as a function of $\rho_i$ for $T=1.4$ and $1.5$ is shown in Figs.~\ref{fig:a3_a1}(b). $A_3(\rho_i)$ increases with increasing $\rho_i$ for both temperatures.

Spin-$0$ impurities which bridge between sites occupied by opposite spins lower the overall energy of the system by a greater amount than impurities surrounded entirely by spins of the same sign. Hence it is the presence of spin-$0$ impurities at the cluster boundary (rather than inside the cluster) which has the greatest impact on the nucleation barrier. With static impurities, the average impurity density per unit length of cluster circumference is independent of both temperature and cluster size, and hence we can capture their impact on the interfacial free energy via this simple functional form. 

Now we compare the critical nucleus size and the barrier height obtained from our modified free energy expression and from the simulation. From Eq.~\ref{eq:f_lambda} it is easy to show that the critical nucleus size may be written as
\be
\label{eq:lambda_c}
\lambda_c(\rho_i)=\bigg[\frac{A_1(\rho_i)+\sqrt{A_1^2(\rho_i)+32 h A_2}}{8h}\bigg]^2.
\ee
The variation of $\lambda_c$ and $F(\lambda_c, \rho_i)$ for different $\rho_i$ obtained from theory and simulations are shown in Figs.~\ref{fig:lambdac_fc}(a) and \ref{fig:lambdac_fc}(b) respectively for two different $T$. The results agree satisfactorily. As would be expected from the reduced interfacial free energy, increased density of impurities results in a smaller critical nucleus size and barrier height. 

\section{Nucleation rates}
\label{nucl-rate}
\begin{figure}[t!]
\includegraphics[width=\columnwidth]{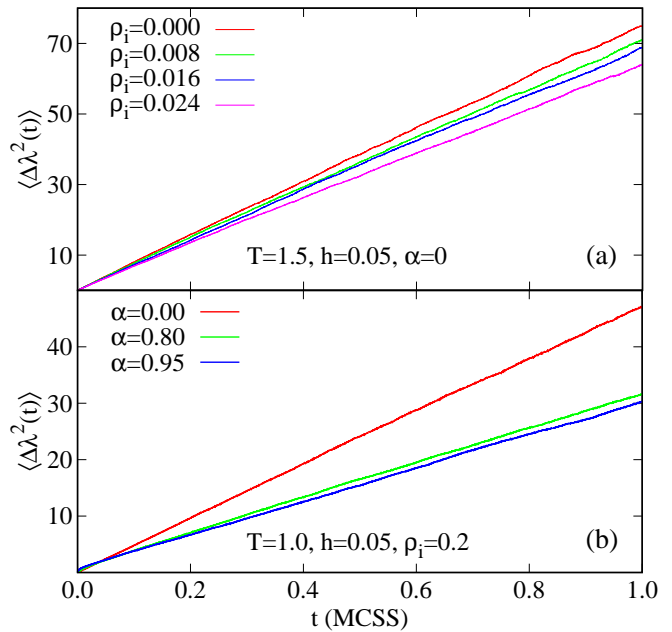}
\caption{Variation of $\langle{\Delta \lambda}^2(t)\rangle$ as a function of time $t$ with (a) static impurities for different densities and (b) dynamic impurities for different $\alpha$.}
\label{fig:diff_comb}
\end{figure}
We have calculated nucleation rates using both BD theory, and the FFS technique (which is independent of CNT assumptions). The expression of the nucleation rate using FFS is shown in Eq.~(\ref{eq:ffs_rate}).
We compare these nucleation rates obtained from  FFS with  nucleation rate estimates from Becker Doring theory using our modified expression for $F(\lambda_c,\rho_i)$ (see Eq.~(\ref{eq:f_lambda_rho}) and Eq.~(\ref{eq:lambda_c})) which incorporates the effects of impurities. The nucleation rate may be written as
\be
I^{BD}=D_c\Gamma e^{-\frac{F(\lambda_c,\rho_i)}{k_BT}},
\label{eq:bd_rate}
\ee
where $D_c$ is a diffusion coefficient which captures the rate of addition to the critical nucleus and $\Gamma$ is the Zeldovich factor defined as
\be
\Gamma=\frac{1}{\sqrt{2\pi k_B T}}\bigg[-\frac{\partial^2F(\lambda,\rho_i)}{\partial \lambda^2}\bigg{|}_{\lambda=\lambda_c}\bigg]^{1/2}.
\ee
The quantity $D_c$ is extracted from the mean squared variation in the nucleus size as plotted in Fig.~\ref{fig:diff_comb}(a) with increasing time for different impurities densities. We perform separate Monte Carlo simulations to calculate $D_c$ for different $\rho_i$. The initial configuration at $t=0$ is taken to be the nucleus with critical size. The mean squared variation is defined as
\be
\Delta \lambda^2(t)=[\lambda(t)-\lambda_c]^2.
\ee
We average  $\Delta \lambda^2(t)$ over an ensemble of $10^5$ such trajectories for each line plotted in Fig.~\ref{fig:diff_comb}(a). The diffusion coefficient $D_c$ is defined as
\be
D_c=\frac{\langle \Delta \lambda^2(t)\rangle}{2t},
\ee
where $\langle\rangle$ represents the ensemble average. The slope gradually decreases with increasing $\rho_i$ indicating that the presence of impurities hinders addition of new positive spins to the cluster. Since this term appears in the exponential pre-factor it has little impact on the nucleation rate, such that the reduction in $F(\lambda_c,\rho_i)$ dominates leading to an overall increase in nucleation rate with increasing impurity density. 

\begin{figure}[t!]
\includegraphics[width=\columnwidth]{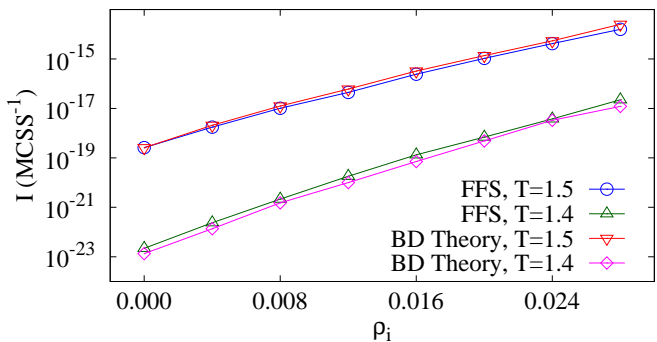}
\caption{Comparison of the nucleation rates obtained from the FFS method and the Becker-Doring theory for $T=1.5$ and $T=1.4$ with $h=0.05$.}
\label{fig:nucl_rate_compare}
\end{figure}
The nucleation rates obtained from the FFS [see Eq.~(\ref{eq:ffs_rate})] and BD theory [see Eq.~(\ref{eq:bd_rate})] are compared in Fig.~\ref{fig:nucl_rate_compare} which shows good agreement. From this we conclude that the impact of static impurities on nucleation rate is well described via classical nucleation theory with only minor extensions required to capture trends with increasing impurity density over the range explored $0\leq\rho_i\leq0.028$. For higher impurity densities we observe multiple overlapping pre-critical nuclei and therefore CNT is not appropriate.

\begin{figure}[t!]
\includegraphics[width=\columnwidth]{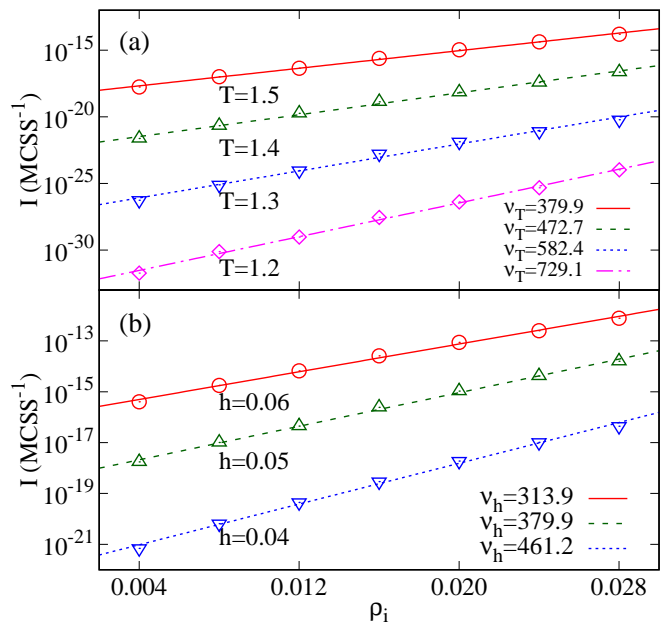}
\caption{Variation of the nucleation rate as a function of $\rho_i$ for different values of the temperatures with fixed external field $h=0.05$.}
\label{fig:nucl_rate}
\end{figure}
We have plotted the nucleation rate $I (MCSS^{-1})$ as a function of $\rho_i$ for different $T$ with fixed $h=0.05$ in Fig.~\ref{fig:nucl_rate}(a). The nucleation rate grows exponentially with increasing impurity density and the exponential growth becomes faster with decreasing  temperature. 
The slope of the best fitted line increases with decreasing $T$ for fixed $h$ as shown in Fig.~\ref{fig:nucl_rate}(a). This indicates that the barrier height decreases more rapidly with increasing $\rho_i$ when the temperature is lowered. The variation of  nucleation rates for different $h$ with fixed $T=1.5$ is shown in Fig.~\ref{fig:nucl_rate}(b). The rates grow exponentially as a function of $\rho_i$ for fixed $T$ and become more steeper with decreasing $h$. We can approximately write the nucleation rate  as $I\approx C\exp{(\nu_T\rho_i)}$ and $I\approx C\exp{(\nu_h\rho_i)}$ for fixed $h$ and $T$ respectively. Both $\nu_T$ and $\nu_h$ increase monotonically with decreasing $T$ and $h$.

\section{Dynamic impurities}
\label{dynamic-imp}

With static spin-$0$ impurities, the reduction of the interfacial free energy is controlled by the overall impurity density. The probability of a impurity being located at the growing $+1$ spin boundary is constant, and independent of temperature. With \textit{dynamic} impurities, the possibility of two limiting regimes arises.
\begin{enumerate}
    \item Low mobility impurities: The boundary of the growing $+1$ spin cluster will expand rapidly compared to the randomly distributed and slow-moving impurities, resulting in a reduction of the interfacial free energy similar to the static case.
    \item High mobility impurities: Sufficiently fast-moving impurities can reach a local equilibrium distribution much faster than the $+1$ spin nucleus can grow or shrink. This implies that a description of the nucleation process in terms of a free energy at each nucleus size remains appropriate.  
\end{enumerate}
Here we apply both the BD and FFS approaches in the high mobility regime and compare to the static regime studied above.

We consider dynamic impurities of fixed density $\rho_i$. Impurity sites may move to the one of the four neighbouring sites via Kawasaki dynamics. The spin $0$ impurity (blue) may exchange with a  spin $+1$ (red) or spin $-1$ (white) neighbouring sites. We do not consider spin exchange dynamics involving only red and white sites. We define the quantity $\alpha$ such that the mobility of impurity particles increases with increasing $\alpha$. The mobility parameter $\alpha$ takes values between $0$ to $1$. In our Monte Carlo simulations we attempt a spin exchange move for a random impurity particle with probability $\alpha$. Attempts to flip the spin of  randomly chosen size with spin $\pm1$ are attempted with probability $1-\alpha$.

Rates are reported to be consistent with the previous section, i.e., based on a unit of time during which, on average, every site on the lattice is subjected to one trial flip. Hence for large $\alpha$, many Kawasaki exchange moves involving impurities are attempted per time unit.

\begin{figure}[t!]
\includegraphics[width=\columnwidth]{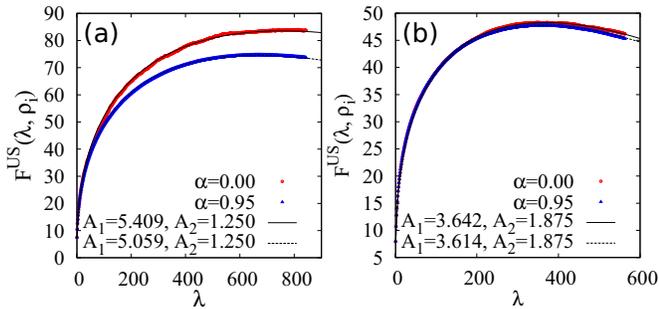}
\caption{Free energy barrier to nucleation with both static ($\alpha=0$) and high mobility ($\alpha=0.95$) impurities at (a) $T=1.0$ and (b) $T=1.5$ with $h=0.05$ and $\rho_i=0.020$. The critical nucleus size at $T=1.0$ for $\alpha=0.0$ and $\alpha=0.95$ are respectively $756$ and $665$.}
\label{fig:dynamic_free_T1.0}
\end{figure}
\begin{figure}[t!]
\includegraphics[width=\columnwidth]{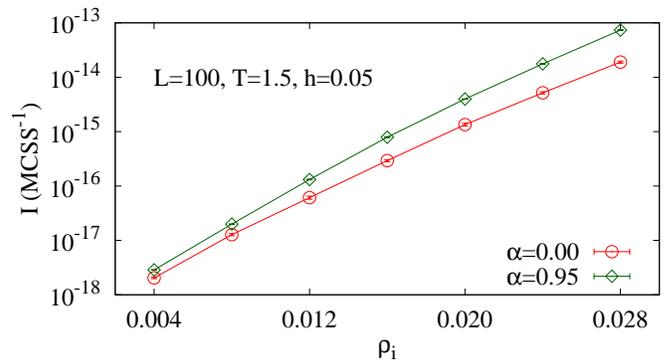}
\caption{Comparison of nucleation rates obtained via FFS for static ($\alpha=0$) and high mobility ($\alpha=0.95$) impurities.}
\label{fig:dynamic_ffs_T1.5}
\end{figure}

We have plotted the free energy barrier to nucleation  both for static ($\alpha=0$) and high mobility ($\alpha=0.95$) impurities with density  $\rho_i=0.020$ for different values of $\alpha$. The free energy plots for  $T=1.0$ and $T=1.5$ are shown in Fig.~\ref{fig:dynamic_free_T1.0}(a) and \ref{fig:dynamic_free_T1.0}(b) respectively.  We observe that the barrier height decreases significantly for the nonzero value of $\alpha$ compared to the $\alpha=0$ case when the temperature is low ($T=1.0$). However, the critical nucleus size is approximately the same. By contrast, we see that the barrier height and critical nucleus size of the free energy do not change as dramatically between $\alpha=0$ and $\alpha=0.95$ at the higher temperature $T=1.5$.

Nucleation rates $I$ as a function of $\rho_i$ obtained using FFS for static and high mobility impurities at $T=1.5$ and $h=0.05$ are shown in Fig.~\ref{fig:dynamic_ffs_T1.5}. We see that the rates obtained using FFS for different $\alpha$ are similar (i.e. within one order of magnitude) for every $\rho_i$ at this high temperature, but there is an indication of  high mobility impurities increasing the nucleation rate.  This is consistent with the observation of marginally lower barrier heights for high $\alpha$ at $T=1.5$ in Fig.~\ref{fig:dynamic_free_T1.0}(b).

Based on the free energies in \ref{fig:dynamic_free_T1.0}(a), we expect a much more dramatic impact of high mobility impurities at the lower temperature of $T=1.0$. The FFS method becomes extremely inefficient at this temperature due to a very low density of spin $+1$ sites in the metastable phase.  Computing a sufficiently accurate flux through $\lambda_0$ is intractable in this case. 

We have, however, calculated the nucleation rate at $T=1.0$ using the Becker Doring expression. The rates are $4.005\times 10^{-38} MCSS^{-1}$ and $2.180\times 10^{-34} MCSS^{-1}$ for $\alpha=0$ and $\alpha=0.95$ respectively. The mobility of impurities hence enhances the nucleation rate by several orders of magnitude.  The mean squared deviation $\langle\Delta\lambda^2(t) \rangle$ for dynamic impurities with different $\alpha$ at $T=1.0$ is shown in Fig.~\ref{fig:diff_comb}(b). We see that the slope of the plot reduces for $\alpha\neq0$ implying the reduction of diffusion coefficient $D_c$ for dynamic impurities.

The BD rates at $T=1.5$ are $1.128\times 10^{-15} MCSS^{-1}$ and $1.930\times 10^{-15} MCSS^{-1}$ for $\alpha=0.95$ and $\alpha=0.0$ respectively. The corresponding critical cluster sizes are $\lambda_c=368$ and $\lambda_c=363$ for dynamic and static case respectively.

\begin{figure}[t!]
\includegraphics[width=\columnwidth]{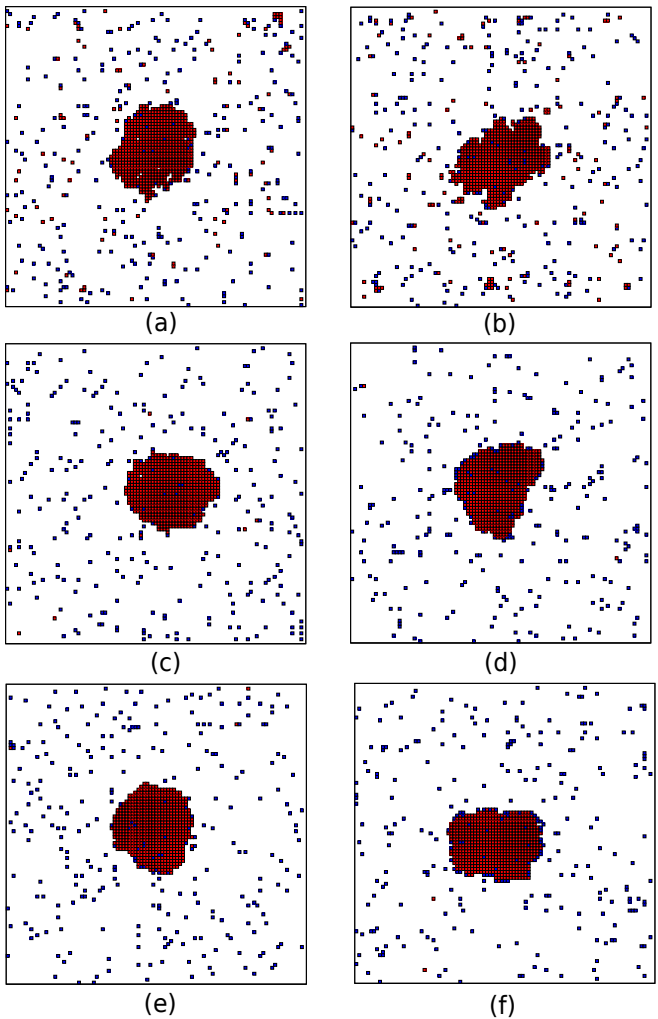}
\caption{Snapshot of the system after equilibration with static (left column) and dynamic impurities (right column) at (a-b) $T=1.5$, (c-d) $T=1.0$ and (e-f) $T=0.9$ with fixed $h=0.05$ and $\rho_i=0.020$. The value of $\alpha$ is $0.95$ for the dynamic impurities.}
\label{fig:dynamic_snap}
\end{figure}

The role of high mobility impurities at high versus low temperature for three different temperatures with fixed $h=0.05$ is illustrated in Fig.~\ref{fig:dynamic_snap}(a-f). Fig.~\ref{fig:dynamic_snap}(a-b), Fig.~\ref{fig:dynamic_snap}(c-d) and Fig.~\ref{fig:dynamic_snap}(e-f) are the snapshots of typical configuration after equilibrium at $T=1.5$, $T=1.0$ and $T=0.9$ respectively for both static and dynamic impurities. For the purposes of this illustration we set the mobility parameter $\alpha=0.95$ prioritising the spin exchange dynamics for the impurity particles. Initially the impurities are placed randomly on the lattice. We create a circular nucleus of red (spin $+1$) particles at the centre of the lattice of size $\lambda=600$ and evolve the system such that the cluster size remain confined within a range $[\lambda-10, \lambda+10]$. We see that the impurity particles preferentially occupy the boundary positions of the nucleus at low temperatures. This happens because the presence of the impurity particles at the boundaries reduces the surface free energy. However, at high temperatures the phenomenon of preferential occupancy of the blue particles at the boundary positions is entropically unfavourable leading to a less dramatic enhancement of the nucleation rate. 

\begin{figure}[t!]
\includegraphics[width=\columnwidth]{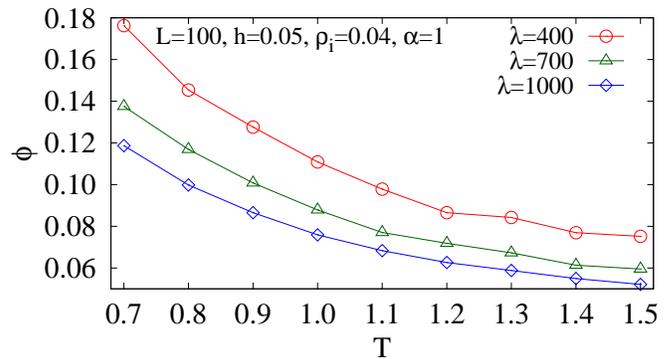}
\caption{Fraction of the blue particles present at the boundary of the nucleus with increasing the temperature for different size of the initial nucleus $\lambda_i$.}
\label{fig:blue_boundary}
\end{figure}

To quantify this observation, we plot the equilibrium fraction $\phi$ of the impurity particles present at the boundary of a static nucleus as a function of $T$ for different sizes of the nucleus as shown in Fig.~\ref{fig:blue_boundary}. The quantity $\phi$ is defined as
\be
\phi=\frac{N_b}{N_r+N_b},
\ee
where $N_r$ is number of red particles connected to the nucleus through the nearest neighbour sites and $N_b$ is number of blue particles present at the boundary of the nucleus and also connected to the nucleus through the nearest neighbour sites. We set the mobility parameter $\alpha=1$ such that only spin exchange dynamics are performed for the impurity particles. The boundary particles are identified by measuring the distance $r$ of the blue particle from the centre of mass of the nucleus. If $r\geq 0.7 R$, we count the blue particle as boundary blue particle, where $R$ is mean radius of the nucleus. The quantity $\phi$ increases with decreasing $T$ indicating the adsorption phenomenon of impurity particles at the boundary of the nucleus. This adsorption phenomenon is absent at high temperatures.


\section{Conclusion}
\label{conclusion}

We have studied nucleation rates in the two dimensional Ising model in the presence of randomly placed impurities on a square lattice using two independent methods which are BD theory and FFS. The rates obtained using both methods agree satisfactorily for different impurity densities upto $\rho_i=0.028$, i.e., when the only $2.8 \%$ of the sites are occupied by the impurities. The nucleation rate decreases with increasing the impurity density. We have also studied the effect on the nucleation rates when the impurity particles are dynamic. The dynamics of the impurity particles accelerate the nucleation  when the temperature is low. However, the impurity dynamics does not play as significant a role when the temperature is high.

We have not explored the regime of intermediate impurity mobility. Here one would not expect the distribution of impurities to equilibrium on a timescale more rapid than that on which spin $+1$ cluster grow or shrink. This would invalidate a free-energy based description, but could be amenable to a description based on BD theory if growth/shrinkage rates could be evaluated directly. Rates in this regime can possibly be probed via FFS, however in the high temperature regime accessible to this method we expect little impact of mobility on nucleation rate due to the less-dramatic difference between the two limiting cases of zero and high mobility. 

The study of the nucleation in the presence of random impurities on triangular lattice and three dimensional cubic lattice is an open area for future research. It would be interesting to investigate how the different dynamics of the impurity particles in other lattice geometries impact the nucleation rate.

\section*{Acknowledgement}

We thank I. J. Ford for helpful discussions. We acknowledge the support from the EPSRC Programme Grant (Grant EP/R018820/1) which funds the Crystallization in the Real World consortium. In addition, we gratefully acknowledge the use of the computational facilities provided by the University of Warwick Scientific Computing Research Technology Platform.

\section*{Data Availability}

Data associated with this manuscript is available via the University of Warwick Research Archive Portal at \url{http://wrap.warwick.ac.uk/155054/}.

%

\end{document}